\documentclass[twoside]{article}
\usepackage{fleqn,espcrc2,psfig}

\newcommand{\MM}{\hspace{-0.24cm} }

\title{Static three quark potential in the quenched lattice QCD}

\author{H. Matsufuru\address[RCNP]{%
     Research Center for Nuclear Physics, Osaka University,
     Mihogaoka 10-1, Ibaraki  567-0047, Japan}%
     \thanks{talk presented by H.Matsufuru at Lattice 2000,
     Bangalore, India.},
  Y. Nemoto\address{Yukawa Institute for Theoretical Physics,
     Kyoto University, Kitashirakawa, Kyoto 606-8502, Japan},
  H. Suganuma\address{Tokyo Institute of Technology,
     2-12-1 Ohokayama, Meguro, Tokyo 152-8551, Japan},
  T.T. Takahashi\addressmark[RCNP]
  and T. Umeda\address{Department of Physics, Hiroshima University,
  1-3-1 Kagamiyama, Higashihiroshima, 739-8526, Japan}}

\begin{document}

\begin{abstract}
We study the static three quark potential using lattice QCD
simulation.
At zero temperature, the three quark potential is extracted from the
baryonic Wilson loop
and fitted to the phenomenological form,
the sum of the Coulomb term, linearly rising term and the constant.
We compare two types of the linear term, ``$Y$'' and ``$\Delta$'' 
type, and find that the string tension almost coincide with
$\sigma_{Q\bar{Q}}$ in the former case. 
We also investigate the three quark potential at finite temperature
using the Polyakov loop correlation.
In the deconfined phase, the screened potential
is observed similarly to the case of $Q\bar{Q}$ system.
\end{abstract}

\maketitle

\section{Introduction}

Determination of the static quark-antiquark ($Q$-$\bar{Q}$) potential
has been one of
most vital subjects of the lattice QCD simulation \cite{Sch00}.
On the other hand, the status of the three quark (3$Q$) potential is
much less settled, in spite of its importance in the baryon spectroscopy.
In this paper, we present our recent result of lattice calculation of
the static three quark potential \cite{Rabbit,Rabbit2}.

It has been widely used the static $Q$-$\bar{Q}$ potential 
expressed as the sum of the short range Coulomb term and long range
confining linear term.
In the case of the 3$Q$ potential,
two distinct treatments of the linear term are frequently applied.
In this paper they are denoted as ``$\Delta$-type'' and ``$Y$-type'' forms.
The former expresses the linear potential as the sum of two-body
potentials, which are proportional to the distance between each
pair of quarks.
The latter treats the potential with the string linking
three quarks at the point with which the total length of the string
takes the minimum length, $L_{min}$.
$Y$-type ansatz has been used in the phenomenological studies
with the string tension $\sigma_{3Q}\simeq \sigma_{Q\bar{Q}}$,
and successfully describes the gross feature of the baryon spectrum
\cite{CI86}.
$\Delta$-type ansatz is, however, also used in some model
calculations with the string tension $\sigma_{\Delta}$ effectively
around half of $\sigma_{Q\bar{Q}}$.
The lattice simulations have been done mainly in 1980's
\cite{SW84,EFLM87,TES87},
and there is one recent preliminary work \cite{Bali00}.
Among them, several authors interpreted their results as the
support of the $\Delta$-type picture \cite{SW84,TES87,Bali00}.
This embarrassing situation is mainly caused by the insufficient
extraction of the ground state contribution from the baryonic
Wilson loop.
In nineties, calculation of $Q$-$\bar{Q}$ potential became much
reliable by making use of the smearing technique which enhances the
ground state contribution to the Wilson loop \cite{APE87}.
This procedure is directly applicable to the three quark system,
and is expected to enable us reliable extraction of the
ground state contribution.

Our first goal of this work is an extraction of reliable static
3$Q$ potential using the smearing technique.
With the lattice result, we discuss preferable description
of the potential.
These subjects are described in the next section.

Another important subject on the 3$Q$ potential is the
thermal effect on it at finite temperature.
Again the lattice computation is much retarded compared with
the $Q$-$\bar{Q}$ potential.
We present our preliminary result at $T>T_c$ in Section 3.

\section{Zero temperature result}

\begin{figure}[tb]
\vspace*{0.1cm}
\centerline{\psfig{figure=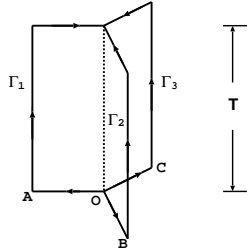,width=3.2cm}}
\vspace{-1.1cm}
\caption{
Baryonic Wilson loop.}
\label{fig:bloop}
\vspace{-0.6cm}
\end{figure}

The three quark potential is extracted from the baryonic Wilson loop,
\begin{equation}
 W_{3Q}(\vec{r}_1,\vec{r}_2,\vec{r}_3,T)
   \equiv \frac{1}{3!}\epsilon_{abc}\epsilon_{a'b'c'}
        U_1^{aa'} U_2^{bb'} U_3^{cc'},
\end{equation}
\begin{equation}
 U_k \equiv P \exp \{ ig \int_{\Gamma_k} dx_{\mu}A_{\mu}(x) \},
\end{equation}
where $P$ denotes the path-ordered product along the path $\Gamma_k$
($k=$1, 2, 3) as shown in Figure~\ref{fig:bloop}.
The position of the contraction of three lines $U_k$ can be chosen
arbitrarily, and the result should not depend on this point.
We choose the contracting point as the spatial origin, and locate
three static quarks on each three spatial axis.
The same constellation is averaged making use of the translational
invariance.

The numerical simulation is done on the lattice of size $12^3 \times 24$,
with the standard Wilson gauge action at $\beta=5.7$ in the quenched
approximation.
The lattice spacing is determined as $a \simeq 0.19$ fm by setting
the string tension of $Q$-$\bar{Q}$ potential to be $\sigma=0.89$ GeV/fm.
The (baryonic) Wilson loop is measured on 210 configurations
with applying the smearing technique to enhance the ground state
contribution.
The potentials are extracted from the (baryonic) Wilson loop by the
single exponential fit in appropriate fit regions, where the Wilson loop is
dominated by the ground state contribution \cite{Rabbit}.

Let us start with the $Q$-$\bar{Q}$ potential.
It is widely accepted that the lattice result of the static $Q$-$\bar{Q}$
potential is well described by the form
\begin{equation}
V(r) = C   - \frac{A}{r}  + \sigma r .
\label{eq:potential_fit}
\end{equation}
In Table~\ref{tab:fit}, we list the results of the fit
in three kinds of fit ranges:
on-axis fit, fits with off-axis data in the whole $r$ region
and at $r\geq 2$.
Since our lattice spacing is rather coarse, the fit with the off-axis
data may suffer from the rotational symmetry breaking effect.
In this sense, the fit with the on-axis data is preferable
to determine the parameters.
For the 3$Q$ potential, however, we inevitably treat the
off-axis constellations, and then it would be meaningful to
examine the typical size of the rotational
symmetry breaking effect in the $Q$-$\bar{Q}$ system.
For this purpose, we carry out the fit including the off-axis data
with two fit ranges.
The fit using all data (all $r$ in Table~\ref{tab:fit}) results
in large $\chi^2$ value, and this is mainly caused by the result at the
short distance region where the statistical error is smallest and
the cutoff effect is most severe.
We alternatively fit at $r\geq 2$, which is more appropriate for
comparison with the result of the 3$Q$ potential.
In this case the fit results in $\chi^2/N_{df} = 3.3$, and
we regard this value as a guide of the rotational symmetry breaking
effect on this lattice.
In both cases, the fit result of the Coulomb coefficient much differs
from the value of on-axis fit than the case of the string tension.

\begin{figure}[tb]
\vspace*{-0.3cm}
\centerline{\psfig{figure=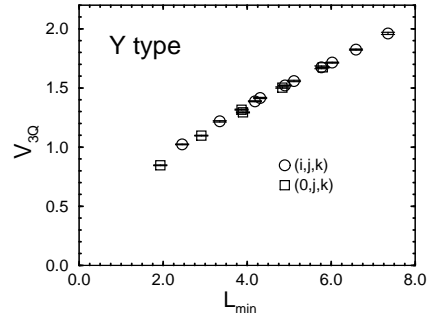,width=6cm}} 
\vspace{-1cm}
\caption{
3$Q$ potential from the viewpoint of $Y$-type ansatz.
$L_{min}$ is the minimum length of the flux linking three quarks.}
\label{fig:pot2}
\vspace{-0.5cm}
\end{figure}

Figure~\ref{fig:pot2} shows the lattice result of the 3$Q$
potential from the viewpoint of the $Y$-type ansatz.
The horizontal axis, $L_{min}$, is the minimum length of the
flux lines linking three quarks.
Three flux connecting the junction with quarks take the angles $2\pi/3$,
which balances three forces with equal strength,
in the case that all three angles of the ``quark triangle''
are less than $2\pi/3$ \cite{CI86}.
The data are fitted to the form
\begin{eqnarray}
V(\vec{r}_1,\vec{r}_2,\vec{r}_3)
 &=& C_{Y}
  - A_{Y} \left( \frac{1}{r_{12}} + \frac{1}{r_{23}}
                          + \frac{1}{r_{31}} \right) \nonumber \\
 & & + \sigma_{Y} L_{min} .
\end{eqnarray}
The result is listed in Table~\ref{tab:fit}.
This form of the potential well describes the observed result,
and the string tension is almost same as in the case of $Q$-$\bar{Q}$
potential.
The $\chi^2$ of the fit is in the same order as in $Q$-$\bar{Q}$
with fit range $r\geq 2$.
Considering the rotational symmetry violation effect at this $\beta$,
this seems to be a consistent result.

Alternative ansatz, $\Delta$-type, is represented with the fitting function
\begin{eqnarray}
V(\vec{r}_1,\vec{r}_2,\vec{r}_3)
 &=& C_{\Delta}
  - A_{\Delta} \left( \frac{1}{r_{12}} + \frac{1}{r_{23}}
                          + \frac{1}{r_{31}} \right) \nonumber \\
 & &  + \sigma_{\Delta} ( r_{12} + r_{23} + r_{31} ),
\end{eqnarray}
and the obtained parameters are also listed in Table~\ref{tab:fit}.
As the result of fit, we find that the string tension is much smaller
than the $Y$-type fit, although the constant and the coefficient
of the Coulomb term are almost the same as the $Y$-type results.
In the fit, $\sigma_{\Delta}$ is slightly larger than the half of
$\sigma_Y$.
Geometrically,
$2 L_{min} \simeq L_{\Delta}=r_{12}+r_{23}+r_{31}$ roughly holds
in most cases.
This explains that $\sigma_{\Delta}\simeq\frac{1}{2}\sigma_{Q\bar{Q}}$.
There is another possible physical explanation
for $\sigma_{\Delta}\simeq \frac{1}{2}\sigma_{Q\bar{Q}}$
using the $SU(3)_c$ Casimir scaling.
For the Coulomb coefficient, one gluon exchange brings
the color factor $1/2$ in the perturbation theory
as the ratio of the $SU(3)_c$ Casimir factor.
The linear term is however nonperturbative and then there is
no clear explanation to keep the $SU(3)_c$ Casimir scaling
beyond the perturbation.
In addition, the $\chi^2$ of the $\Delta$-type fit is much larger
than the case of the $Y$-type fit, where the goodness of
the fit is comparable with $Q$-$\bar{Q}$ potential case with $r\geq 2$.
Although the present simulation may suffer from rather large
finite $a$ effect, our result shows that the $Y$-type ansatz
is more preferable with the lattice data.

\begin{table}[tb]
\begin{center}
\begin{tabular}{ccccc}
\hline\hline
  & $C$ & $A$ & $\sigma$ & \MM\MM$\chi^2$/$N_{df}$\MM \\
\hline
\MM 3$Q$({\small $Y$})\MM 
 & 0.914(20)\MM& 0.1316(62)\MM& 0.1528(20)\MM& 4.0\MM \\
\hline
\MM 3$Q$({\small $\Delta$})\MM 
 & 0.934(20)\MM& 0.1405(60)\MM& 0.0858(15)\MM& 11.\MM \\
\hline
\MM $Q$-$\bar{Q}$ \\
\MM on-axis  \MM& 0.629(16)\MM& 0.2793(12)\MM& 0.1629(47)\MM& 0.59\MM \\
\MM  all $r$ \MM& 0.686( 7)\MM& 0.3351(55)\MM& 0.1545(20)\MM& 16. \MM \\
\MM $r\geq 2$\MM& 0.696(24)\MM& 0.395 (35)\MM& 0.1561(38)\MM& 3.3 \MM \\
\hline\hline
\end{tabular}
\end{center}
\caption{Fit results of the $Q$-$\bar{Q}$ and the 3$Q$ potentials.}
\label{tab:fit}
\vspace{-0.4cm}
\end{table}

Our conclusion at this stage is the following.
$Y$-type ansatz is more preferable for the lattice result
although $\Delta$-type ansatz is not completely excluded with
present systematic uncertainty.
Adopting $Y$-type picture, we have found a universal feature of the
string tension as $\sigma_{3Q}\simeq \sigma_{Q\bar{Q}}$.
To confirm this result and investigate the 3$Q$ potential in more
detail, simulations at larger $\beta$'s are in progress.
Preliminary result on the $16^3\times 32$ lattice at
$\beta=6.0$ is consistent with the result at $\beta=5.7$.

\section{Finite temperature result}

Now we turn on the temperature and observe how the potential changes
with the thermal effect \cite{Rabbit2}.
In this paper, we focus on the static 3$Q$ potential in the
deconfined phase.
Above the critical temperature $T_c$, the quarks and the gluons are
expected to be liberated 
and form a plasma state, in which the potential is expected to be screened.
Then one expects that the screened potential takes the Yukawa potential
like as the $Q$-$\bar{Q}$ potential \cite{Gao90}, and genuine three-body
potential (in $Y$-type picture) would disappear.

On the finite temperature lattice, the static potential is extracted
from the Polyakov loop correlation,
\begin{eqnarray}
 P_3(\vec{r}_1,\vec{r}_2,\vec{r}_3)
  = \langle P(\vec{r}_1)P(\vec{r}_2)P(\vec{r}_3) \rangle
 \nonumber \\
  = c \cdot \exp(-V_{3Q}(\vec{r}_1,\vec{r}_2,\vec{r}_3) N_{\tau}),
\end{eqnarray}
where the Polyakov loop $P(\vec{x})$ is defined as 
$P(\vec{x}) = \mbox{\rm Tr} \prod_{t=0}^{N_{\tau}-1} U_4(\vec{x},t)$.
The $Q$-$\bar{Q}$ potential is also measured in similar way.

To simulate the finite temperature system, we adopt the anisotropic
lattice \cite{Kar82}.
An advantage of the anisotropic lattice for our study is that it
enables us to work out several values of temperature without varying
the lattice cutoff.
As the gauge field action, we use the Symanzik improved action
at the tree level \cite{Sym83}.

The numerical calculation is done on the lattices of sizes
$16^2\times 24\times N_t$ at $\beta=4.56$ and $\gamma_G=3.45$,
in the quenched approximation.
At $N_t=96$, which is realized as the zero temperature lattice,
we determine the renormalized anisotropy as $\xi=3.95(2)$ 
from the Wilson loops in the spatial-spatial and temporal-spatial planes.
The static $Q$-$\bar{Q}$ potential at $T=0$ gives the cutoffs
$a^{-1}_{\sigma}=1.61(1)$ GeV and  $a^{-1}_{\tau}=6.36(5)$ GeV.
The critical temperature is estimated by the rapid change of the
Polyakov loop and its susceptibility, and is found between
$N_t=24$ and $25$, and slightly near to $N_t=24$.
For the investigation of the static potentials in the deconfined
phase, we generate 60 configurations at $N_t=20$, which
roughly corresponds to $T\simeq 1.22 T_c$.

\begin{figure}[tb]
\centerline{\psfig{figure=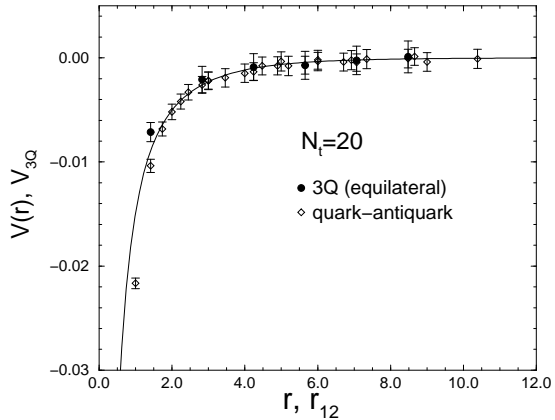,width=7.8cm}}
\vspace{-1.2cm}
\caption{
$Q$-$\bar{Q}$ (open symbols) and 3$Q$ (filled) potentials at
$N_t=20$ ($T>T_c$).
The 3$Q$ potential is divided by 3.
}
\label{fig:FT2}
\vspace{-0.5cm}
\end{figure}

Figure~\ref{fig:FT2} shows the $Q$-$\bar{Q}$ potential and the 3$Q$
potential in the deconfined phase.
At the long distance, The Polyakov loop correlators approach to the
multiples of the expectation value of single Polyakov loop:
$P_2\rightarrow \langle P \rangle \langle P \rangle^*$ and 
$P_3\rightarrow \langle P \rangle^3$.
These contributions are subtracted from the potentials.
The $Q$-$\bar{Q}$ potential above $T_c$ is known to behave as the
Yukawa-type potential \cite{Gao90}.
In present calculation, the Debye screening mass of the $Q$-$\bar{Q}$
potential strongly depends on the fit range and is estimated
around 0.5 -- 1  GeV.
At this $N_t$, we have measured the 3$Q$ potential with only
equilateral constellations.
In Figure~\ref{fig:FT2}, we show the 3$Q$ potential divided by 3,
and find good agreement with the  $Q$-$\bar{Q}$ potential,
\begin{equation}
V_{3Q}(r_{12}\!=\!r_{23}\!=\!r_{31}\!=\!r) \simeq 3 V_{Q\bar{Q}}(r)
\hspace{0.3cm} (T\!>\!T_c).
\end{equation}
Above $T_c$, since the quarks are deconfined, there is no reason to assume
that the combination of quarks are in the color-singlet states.
The question is whether the 3$Q$ potential can be represented
as the sum of two body potential.
Our present result shows that the $V_{3Q}/3$ behaves as almost same as 
the $Q$-$\bar{Q}$ potential at least for the equilateral 3$Q$ constellations.
For more definite conclusion, we need precise results including other
3$Q$ constellations and a systematic comparison between $V_{3Q}$
and $V_{Q\bar{Q}}$ at various $T$.

\bigskip

\noindent
The calculations have been done on NEC SX-4 at Research Center for 
Nuclear Physics, Osaka University and NEC HSP at
INSAM (Institute for Nonlinear Science and Applied Mathematics),
Hiroshima University.


\begin{thebibliography}{99}

\bibitem{Sch00}  % Q-bar{Q} potential review
 K. Schilling, Nucl. Phys. B (Proc. Suppl.) 83-84 (2000) 140,
   and references therein.

\bibitem{Rabbit} % our publication
 T.T. Takahashi, H. Matsufuru, Y. Nemoto and H. Suganuma,
  hep-lat/0006007.

\bibitem{Rabbit2} % our publication
H. Matsufuru, Y. Nemoto, H. Suganuma, T.T. Takahashi and T.Umeda, 
 Proc. of "Quantum Chromodynamics and Color Confinement", in press.

\bibitem{CI86} %  model calculation
 S. Capstick and N. Isgur, Phys. Rev. D 34 (1986) 2809.

\bibitem{SW84} % 3Q potential (lattice)
 R. Sommer and J. Wosiek, Phys. Lett. B 149 (1984) 497;
                          Nucl. Phys. B 267 (1986) 531.

\bibitem{EFLM87}  % 3Q potential (lattice)
 J. Kamesberger, G. Eder, M.E. Faber, H. Leeb and H. Markum, 
   Proc. of ``Few-Body Problems in Particle, Nuclear, Atomic and Molecular
   Physics''(1987) 529. % in Fontevraud 

\bibitem{TES87} % 3Q potential (lattice)
 H.B. Thacker, E. Eichten and J.C. Sexton,
   Nucl. Phys. B (Proc. Suppl.) 4 (1988) 234.

\bibitem{Bali00} % 3Q potential (lattice)
 G.S. Bali, hep-lat/0001312.

\bibitem{APE87} % Smearing technique
 APE Collaboration (M.Albanese et. al.),
   Phys. Lett. B 192 (1987) 163.

\bibitem{Gao90}  % QQ potential at FT
 M. Gao, Phys. Rev. D 41 (1990) 626.

\bibitem{Kar82} %   Anisotropic lattice
 F. Karsch, Nucl. Phys. B 205 (1982) 285.

\bibitem{Sym83} %  Symanzik action
 K. Symanzik, Nucl. Phys. B226 (1983) 187.

\end{thebibliography}
\end{document}